
It is well known that quantum fluctuations of massless scalar and tensor
(gravitational) fields are very much amplified at inflationary stage and
create considerable density inhomogeneities [1-4] or relic gravitational
waves [5-12]. This is closely related to the fact that these fields are not
conformally invariant even though they are massless. Conformal invariance of
a massless scalar field $\phi$ can be only achieved by the nonminimal
coupling to the curvature scalar $R\phi^2 /6$ and in this case quantum
fluctuations are not amplified in the Robertson-Walker background.
Gravitational waves are also conformally noninvariant in the standard General
Relativity [13]. The amplification of the quantum fluctuations in cosmological
conditions can be understood as particle production by external gravitational
field. Since the Robertson-Walker metric is known to be conformally flat
the background gravitational field does not produce particles if the
underlying theory is conformally invariant [14]. In particular electromagnetic
waves are not produced in such conditions since the classical electrodynamics
is conformally invariant in the limit of vanishing masses of fermions. Quantum
corrections however are known to break the conformal invariance. There are
three
possible sources of its breaking: nonzero masses of charged particles,
absence of conformal invariance for (even massless) scalar field as mentioned
above, and quantum conformal anomaly due to the celebrated triangle
diagrams [15]. In what follows we consider the last case because the
other two produce generically much smaller effects.

It was shown in ref. [16] that production of photons
in conformally flat
cosmological background due to the trace anomaly
can be considerable and that the Maxwell equations are modified by the anomaly
in the following way:
$$\partial_\mu F^\mu _\nu + \kappa {\partial_\mu a \over a} F^\mu_\nu =0
\eqno (1)$$
where $a=a(\tau ) $ is the scale factor, $\tau$ is the conformal time,
the metric has the form
$ds^2 =a^2(\tau )(d\tau^2 -dr^2 )$, and the contraction of the indices is made
with the metric tensor of the flat space-time. The numerical coefficient
$\kappa$ in $SU(N)$-gauge theory with $N_f$ number of charged fermions is
equal to
$$\kappa =    {\alpha \over \pi}\left( {11N \over 3} -{2N_f \over 3}\right)
\eqno (2)$$
Here $\alpha $ is the fine structure constant which is to be taken at the
momentum transfer $p$ equal to the Hubble parameter during inflation,
$p=H$. In the asymptotically free theory one would expect
$\alpha \approx 0.02$.

It is convenient to chose the gauge condition $\partial _\mu A^\mu =0$. In
this gauge the time component of the vector potential $A_\tau$ satisfies the
same equation as in the conformally invariant case while the space component
are modified by the extra term in the equations of motion:
$$(\partial ^2_\tau - {1 \over a^2} \partial ^2_l +
\kappa {a' \over a} \partial_\tau )A_j (\tau, r)=0
\eqno (3)$$
where prime means differentiation with respect to conformal time $\tau$ and we
put $A_\tau =0$ since this component is not amplified by the Robertson-Walker
background.

Quantization of the electromagnetic field in this background metric with the
account of the conformal invariance breaking can be made with the usual
decomposition
$$A_j (\tau , r) =\int {\,d^3k \over (2\pi)^{3/2} (2\omega)^{1/2}}
[c_j (k) A (\tau , k) e^{-ikr} + h.c. ] \eqno (4)$$
where operators $c_j$ satisfy
$$\langle c_j (k) c^{\star l}\rangle _{vac} =
\delta (k' -k) (\delta_j^l -k_jk^l /k^2) \eqno (5)$$
The C-function $A(\tau , k)$ satisfies the equation
$$A'' +k^2 A +\kappa {a' \over a} A' =0 \eqno (6)$$

Note that the amplitudes of massless minimally coupled scalar field and of
gravitational wave satisfy the same equation with $\kappa =2$.

Equation (6) is solved as
$$A( k, \tau ) = (k\tau )^{1/2 -\kappa} [C_1 J_{\kappa - 1/2} (k\tau ) +
C_2 J_{1/2 -\kappa} (k \tau ) ] \eqno (8)$$
The coefficients $C_1$ and $C_2$ can be found from the conditions that
$A(k, \tau )$ tends to $\exp (i\omega t)$ in the limit of vanishing Hubble
parameter. In the De Sitter stage $\tau \approx (1/H -t)$,
as $H\rightarrow 0$,
where $t$ is the physical time. Using this equation and the asymptotic
expressions for the Bessel functions  (as $k\tau \rightarrow \infty $) we find
$$C_1 = {\sqrt {2\pi } \over 1-\exp (2i\pi\nu ) } \left( {H \over k }\right)^
{\kappa /2} \exp ({ik\over H} +{i\pi \over 4} -{i\pi \nu \over 2} )$$
$$c_2 =-\exp (-i\pi \nu ) \eqno (8)$$
where $\nu =(\kappa +1) /2$.

Near the end of inflation that is for $\tau \rightarrow 0, $ $A(k, \tau )$
is given
by the expression
$$|A(k, \tau )| \rightarrow {2^\nu \sqrt {2\pi}
\over |1-\exp (2i\pi \nu)|\Gamma (\nu +1)} \eqno (9)$$
Using this expression and the relation between the physical and conformal
momenta $p=a(t)k$ we can evaluate the energy
density of electromagnetic field generated
during inflation at the moment when its wave reenters the horizon as
$$F_{\mu \nu} ^2 \approx (Hl)^\kappa /l^4  \eqno (10) $$
If $\kappa \ll 1$ the corresponding field is negligibly small but for
$\kappa =O(1)$ the amplitude of the magnetic field can be large enough to
seed the observed magnetic fields in galaxies. Indeed it is known from
observations that the strength of
the galactic magnetic fields is
of the order of $B=10^{-6}G$ and correspondingly its
energy density is close to that
of the electromagnetic background radiation.
The field given by eq. (10) is about
$10^{40}$ below the necessary value if $(Hl)^\kappa  =O(1)$. However the
factor $Hl$ is extremely large. For $H=10^{12}$ GeV and $l=L_{gal}$ it is
about $10^{46}$. So for $\kappa =O(1)$ the magnetic field generated during
inflationary stage can be large enough to create the observed fields in
galaxies even without
the dynamo amplification [17,18]. The latter may amplify
the seed magnetic field by about 10 orders of  magnitude permitting a
slightly smaller value of $\kappa$.
In contrast to magnetic field the electric field generated during inflationary
stage not only is not amplified but vice versa is dumped down due to the
large conductivity of the primeval plasma. One may argue that the conductivity
is nonzero and large already during the De Sitter stage because of the thermal
background with the temperature $H/2\pi$ [19]. This background is associated
with the horizon at $1/H$. However the particle polarization which could damp
the electric field cannot compete with the universe expansion and hence is not
able to screen the field. Electric field should be screened during the Friedman
stage after the inflaton energy density is transformed into energy density
of real particles in the
cosmic plasma. The characteristic time scale of this process is of
the order of the horizon and so one may expect generation of chaotic electric
currents on cosmologically large scales. These chaotic electric currents
may in turn generate cosmic magnetic fields.

Note that there are several other proposals in the literature for magnetic
field
generation in a theory with broken conformal invariance. In ref [20] it was
assumed that there exists the nonminimal coupling of the curvature scalar
to electromagnetic filed of the form $\xi R A_\mu ^2$ so that not only
conformal but also gauge invariance of electromagnetism is broken. Another
proposal [21] is the coupling of the electromagnetic field strength tensor
to the dilaton field $\exp (\phi) F^2_{\mu\nu}$. In both these cases it was
argued that it is possible to generate the seed magnetic field of the
appropriate amplitude during inflation. Our model does not demand any
extra coupling in addition to the standard electrodynamics but to get a large
enough field the number of charged particles with $m<H$ should be about 30.
Reverting the arguments one can put a bound on $\kappa$ from the absence of
too strong cosmic electromagnetic fields.

I am grateful to B. Ratra, Sun Hong Rhie
and M. Voloshin for critical comments and
to the Institute of Theoretical Physics of the
University of Minnesota where this work was started and to the Department
of Physics at the University of Michigan where it was completed, for the
hospitality.

\bigskip

\centerline {{\bf REFERENCES}}

\item{1} A.H. Guth and S.-Y. Pi, Phys. Rev Lett. {\bf 49}, 1110 (1982)

\item{2} S.W. Hawking, Phys. Lett. {\bf 115B}, 295 (1982)

\item{3} A. A. Starobinsky, Phys. Lett. {\bf 117B}, 175 (1982)

\item{4} J. M. Bardeen, P. J. Steinhardt, and M. S. Turner, Phys. Rev. {\bf
D28}, 679 (1983)

\item{5} A. A. Starobinsky, JETP Letters. {\bf 30}, 682 (1979)

\item{6} V. A. Rubakov, M. V. Sazhin, and A. V. Veryaskin, Phys. Lett. {\bf
115B}, 189 (1982)

\item{7} A. A. Starobinsky, Sov. Astron. Letters {\bf 9}, 302 (1983)

\item{8} R. Fabri and M Pollock, Phys. Lett. {\bf 125B}, 445 (1983)

\item{9} L. F. Abbot and M. B. Wise, Nucl. Phys. {\bf B244}, 541 (1984)

\item{10} A. A. Starobinsky, Sov. Asron. Lett. {\bf 11}, 133 (1985)

\item{11} R. L. Davis, H. M. Hodges, G. F.Smoot, P. J. Steinhardt, and M. S.
Turner, Phys. Rev. Lett. {\bf 69}, 1856 (1992)

\item{12} A. Dolgov and J. Silk, Phys. Rev. (to be published)

\item{13} L. P. Grishchuk, Sov. Phys. JETP. {\bf 40}, 409 (1975)

\item{14} L. Parker, Phys. Rev. Lett. {\bf 21},562 (1968)

\item{15} M. Chanowitz and J. Ellis, Phys. Rev. {\bf D7}, 2490 (1973)

\item{16} A. D. Dolgov, Sov. Phys. JETP {\bf 44}, 223 (1981)

\item{17} E. N. Parker, Astrophys. J.{\bf 163}, 252 (1971)

\item{18} S. T. Vainshtein and A. A. Ruzmaikin,  Sov. Astronomy {\bf 15}, 714
(1972);{\bf 16}, 365 (1972)

\item{19} G. Gibbons and S. Hawking, Phys. Rev. {\bf D15}, 2738 (1977)

\item{20} M. S. Turner and L. M. Widrow, Phys. Rev. {\bf D37}, 2743 (1988)

\item{21} B. Ratra GRP-286/CALT-68-1750; GRP-287/CALT-68-1751
\end